\documentclass[twocolumn,prb,showpacs,preprintnumbers]{revtex4}

\usepackage{graphicx}

\begin{document}

\title{String order in spin liquid phases of spin ladders}

\author{G.\ F\'ath, \"O.\ Legeza, and J.\ S\'olyom }

\address{Research Institute for Solid State Physics and Optics,
       H-1525 Budapest, P.\ O.\ Box 49, Hungary}
\date{26 February 2001}

\begin{abstract}
Two-leg spin ladders have a rich phase diagram if rung, diagonal and plaquette
couplings are allowed for. Among the possible phases there are two 
Haldane-type spin liquid phases without local order parameter, which differ, 
however, in the topology of the short range valence bonds. We show that these
phases can be distinguished numerically by two different string order 
parameters. We also point out that long range string- and dimer orders can
coexist.
\end{abstract}

\preprint{Physical Review B 63, 134403 (2001)}
\maketitle

\section{Introduction}

The properties of spin ladder systems attracted considerable interest in 
the past decade.\cite{review} It was shown\cite{ferro} that even-leg ladder 
models develop a gap in their excitation spectrum if the rung coupling is 
ferromagnetic. This result could be considered as a generalization of 
Haldane's conjucture\cite{haldane} about the existence of a finite gap in 
integer spin Heisenberg chains, since for strong enough ferromagnetic rung
coupling the spin ladder with spin-$S$ on the legs is equivalent to a spin 
chain of spin-$2S$. In this case the ground state is an AKLT like 
state\cite{AKLT} in which short range valence bonds couple spins on 
neighboring rungs. Later it was demonstrated numerically,\cite{antiferro} 
that a gap is generated for antiferromagnetic rung coupling as well, in which 
case the singlet valence bonds are formed
predominantly on the rungs. It may be called a {\em rung singlet} state.
Both are Haldane-type spin liquids in which there are no broken local 
symmetries, and the excitations are coherent magnons with a finite gap.

Similar situation occurs if the two legs of the ladder are coupled
not by rung couplings, but diagonally or by plaquette couplings.\cite{legeza} 
When these couplings compete with each other, non-Haldane spin liquid 
phases, characterized by the absence of coherent magnon excitations and 
spontaneous dimerization,\cite{legeza,nersesyan}may occur due to 
frustration effects. Although the thermodynamic properties of Haldane 
and non-Haldane spin liquids are identical, the correlation functions 
differ due to the difference in the excitations spectrum. The two phases 
could also be distinguished by a local dimer order parameter. 

Beside these spin liquid phases several other phases may occur, like
ferromagnetic or incommensurate phases. When drawing the phase diagram, 
it was thought for some time that the concept of Haldane-like spin liquids 
is unambiguous for spin ladders, and therefore the AKLT-like and rung 
singlet phases are identical not only thermodynamically, but also in the 
sense, that in a large parameter space, where rung, diagonal and plaquette 
couplings are all taken into account, one could go from one to the other 
without any discontinuity. In fact, it was argued\cite{white1} that there 
is a controlled way of changing the parameters gradually and interpolate 
continuously between two limiting cases belonging trivially to one or to 
the other phase. The procedure contains a shift of one of the legs of the 
ladder by one unit, thereby transforming a rung singlet into a diagonally 
situated valence bond, characteristic of the the AKLT-like state, or vice 
versa. However, as was pointed out by Kim {\em et al.}\cite{kim} this shift 
changes the topology of the system drastically. When we start from an open 
ladder, the above transformation deforms the usual straight-end ladder into 
a ``ladder'' whose ends are cut diagonally. Such a deformation has serious 
consequences on the degeneracy of the ground state for open boundary 
condition. In a usual ladder the AKLT state has a fourfold degeneracy due to
the unpaired end spins, while the rung singlet state is unique. On the other
hand, in the diagonally cut ladder the ground state is unique for the AKLT state,
and fourfold degenerate in the rung singlet state. This is best seen in 
Fig.~\ref{fig:topo}(a) and (b), where the typical valence-bond 
configurations of the AKLT-like state and the rung singlet state are shown. 
In Fig.\ \ref{fig:topo}(c) we also show a typical valence bond configuration
of the dimer phase. In this phase the invariance of translation by one
rung is broken spontaneously.

While in the AKLT state the number of valence bonds crossing an arbitrary 
vertical line between rungs is always {\em odd}, in the rung singlet state
it is always {\em even}. This is true not only in this special configuration
but remains valid generally, as long as the range of valence bonds is short
compared to the size of the system.\cite{bonesteel} The two states can be
distinguished by a topological quantum number. Thus for topological reasons 
in a spin ladder the two Haldane-type spin liquid phases are different and 
in the space of couplings they must be separated by a phase transition
hyperplane,\cite{kim,nedelcu} although extended gapless (critical) regions
between them cannot be excluded a priori either. 

Since the topological quantum number is not accessible to direct computation, 
a generalization of the hidden order parameter of the spin-1 
chain\cite{dennijs} was proposed,\cite{kim} that could distinguish between 
the even and odd topological sectors of the Haldane phase. Although
the relationship between the two new string order parameters and the two
types of Haldane phases was made plausible by looking at special 
configurations, it was not possible to calculate their value, since the
bosonization procedure applied to obtain the phase transition lines did 
not distinguish between the two parameters.

The aim of this paper is to do a numerical study of the relationship
between the possible spin liquid phases of the spin ladders and the string
order parameter, using the density matrix renormalization group (DMRG) 
procedure.\cite{dmrg} The setup of the paper is as follows. In Sec.\ II 
we give a short description of the ladder  models and a brief account
of what is known about the possible phases.
The results of our numerical calculations are 
presented in Sec.\ III. Finally Sec.\ IV contains a brief summary.

\section{Gapped phases in ladder models}

Following the notations of Ref.\ \ref{kolezhuk} we will write the most 
general form of the Hamiltonian for isotropic two-leg spin ladder models if 
only spins on the same and neighboring rungs interact, as
\begin{equation}
  {\cal H}=\sum_{i=1}^N h_{i,i+1} ,
  \label{eq:comp1}
\end{equation}
where  
\begin{eqnarray}
    h_{i,i+1} & = & J_{\rm l}\; \vec\sigma_i \cdot \vec\sigma_{i+1} +
  J_{\rm l}'\; \vec\tau_i \cdot \vec\tau_{i+1}   \nonumber  \\
    &  & + \frac{1}{2} J_{\rm r}\; \vec\sigma_i \cdot \vec\tau_i 
    + \frac{1}{2} J_{\rm r}'\; \vec\sigma_{i+1} \cdot \vec\tau_{i+1}  
          \nonumber  \\
    &  & + J_{\rm d}\; \vec\sigma_i \cdot \vec\tau_{i+1} +
       J_{\rm d}'\; \vec\tau_i \cdot \vec\sigma_{i+1}   \nonumber  \\
    &  & + V_{\rm ll}\; (\vec\sigma_i \cdot \vec\sigma_{i+1})
  (\vec\tau_i \cdot \vec\tau_{i+1})  \nonumber  \\
    &  & + V_{\rm rr}\; (\vec\sigma_i \cdot \vec\tau_i)
       (\vec\sigma_{i+1} \cdot \vec\tau_{i+1})   \nonumber \\
    &  & + V_{\rm dd}\; (\vec\sigma_i \cdot \vec\tau_{i+1})
   (\vec\tau_i \cdot \vec\sigma_{i+1})  \,.
\label{eq:plaquett}
\end{eqnarray}
In Eq.\ (\ref{eq:plaquett}) $\sigma_i$ and $\tau_i$ are spin-1/2 operators
corresponding to spins sitting on the two legs of the ladder. 
The schematic plot of the spin couplings in $h_{i,i+1}$ between
the spins on the two legs of the ladder
are shown in Fig. \ref{fig:ham}.

The terms with $J_{\rm l}$ and $J_{\rm l}'$ couple neighbouring spins 
on the same legs, the terms with $J_{\rm r}$ and $J_{\rm r}'$ spins on the same 
rung, while the terms with $J_{\rm d}$ and $J_{\rm d}'$ spins that are situated
diagonally. In addition to these bilinear terms there could be three
biquadratic terms with couplings $V_{\rm ll}$, $V_{\rm rr}$ and $V_{\rm dd}$.

In order to decrease the number of parameters from now on we only consider 
the case when $J_{\rm R}\equiv J_{\rm r} = J_{\rm r}'$,
$J_{\rm L}\equiv J_{\rm l} = J_{\rm l}'$ and $J_{\rm D}\equiv J_{\rm d}
= J_{\rm d}'$. The energy scale will be set by fixing $J_{\rm L}=1$. 

It is known from previous studies\cite{legeza} that all the interleg couplings 
$J_{\rm R},J_{\rm D},\dots,V_{\rm dd}$, when acting alone,
generate a Haldane-like phase. This is an AKLT-like phase with odd topology 
of the number of valence bonds if $J_{\rm R} < 0$, $J_{\rm D} > 0$, 
$V_{\rm ll} > 0$, $V_{\rm rr} < 0$ or $V_{\rm dd} > 0$, 
while for opposite sign of the couplings a rung singlet like state with 
even topology is generated.

As mentioned before, the goal of the present paper is to identify an 
appropriate order parameter that can characterize the spin liquid phases. 
Starting from the string order parameter of the $S=1$ chain proposed by 
den Nijs and Rommelse\cite{dennijs}, it is natural to generalize it to 
spin ladders by defining the order parameter of AKLT-like Haldane state as
\begin{equation}
{\cal O}^{\alpha}_{\rm odd}=- \lim_{|i-j|\to\infty}\left<  S_i^\alpha\exp
\left( i\pi\sum_{l=i+1}^{j-1}S_l^\alpha\right)S_j^\alpha \right>\,
\label{eq:string_odd}
\end{equation}
where $S_i^\alpha$ is the total spin on rung $i$, $S_i^{\alpha}=
\sigma_i^{\alpha}+\tau_i^{\alpha}$, $\alpha=x,y,z$. 

Since in the extreme limits the pure rung singlet state can be related to the 
AKLT state by a shift of one of the legs, the string order parameter
of the Haldane state with even topological quantum number is expected to
be
\begin{equation}
{\cal O}^{\alpha}_{\rm even}=- \lim_{|i-j|\to\infty}\left<  S_i^\alpha\exp
\left( i\pi\sum_{l=i+1}^{j-1}S_l^\alpha\right)S_j^\alpha \right>\,
\label{eq:string_even}
\end{equation}
where $S_i^\alpha$ is now the sum of two diagonally situated spins,
$S_i^{\alpha}=\sigma_i^{\alpha}+\tau_{i+1}^{\alpha}$.\cite{kim,nishi}

In what follows we will calculate the order parameters for the $z$ component 
only, they will be denoted as ${\cal O}_{\rm odd}$, and ${\cal O}_{\rm even}$,
respectively.  We will show, that these two order parameters are mutually
exclusive. ${\cal O}_{\rm odd}$ 
is finite for the AKLT-like Haldane state, 
while ${\cal O}_{\rm even}$ is finite for the rung singlet like Haldane 
phase. Moreover the order parameter goes to zero continuously at the 
transition point when the transition is of second order, while it vanishes 
with a jump in case of first order transitions.

\section{Numerical results}

To investigate this problem we performed numerical calculations applying 
the DMRG method\cite{dmrg} on the model defined by the Hamiltonian in 
Eq.\ (\ref{eq:comp1}). Since the DMRG procedure is more accurate for 
systems with free ends, we consider our ladder model with open boundary 
condition. In order to eliminate the fourfold degeneracy of the ground 
state in the conventional Haldane phase due to the free $S=1/2$ spins 
at the ends of the ladder, we exploited the left-right reflection symmetry 
of the system. Out of the four degenerate ground state we have selected the
one with $S^z_{\rm tot} =0$ and positive parity.
The calculations were done for chains with up to 150 sites keeping 100 
to 200 states per block. The truncation error varied in the range 
$10^{-5}$ to $10^{-7}$. In most of the calculations we used the 
finite-lattice method with two or three iteration cycles. 

In order to reduce the undesirable boundary effects we fixed site $i$ 
in Eqs.\ (\ref{eq:string_odd}) and (\ref{eq:string_even})
at the center of the chain and let site $j$ run through 
half of the chain. The finite size scaling procedure was used to determine 
the thermodynamic limit of the quantitities measured. 

In what follows we present our numerical results for the string order 
parameters for various choices of the coupling constants. Since our aim 
was not to explore the full phase diagram, but rather to check how the 
string order parameters ${\cal O}_{\rm even}$ and ${\cal O}_{\rm odd}$ 
vary as we go from one Haldane-like phase to another, topologically 
different gapped phase, we calculated the string order parameters 
along specially chosen paths connecting points in the parameter space,  
where the model is known to belong to different phases. 

It is worth pointing out that since the ground state wave function and 
therefore the ground state expectation values of operators can be calculated
more accurately than those of excited states, the order parameters provide a 
better way to obtain phase diagrams than investigating the energy spectrum.  

\subsection{Haldane phases due to rung, diagonal or
plaquette couplings alone}

The effect of $J_{\rm R}$ alone was analysed numerically for short systems 
in Ref.\ \ref{nishi}. It was found there that for $J_{\rm R}<0$ the order 
parameter ${\cal O}_{\rm odd}$ is finite, while for $J_{\rm R}>0$ the 
other order parameter ${\cal O}_{\rm even}$ is finite. Our results with the 
DMRG method confirmed these results.

From earlier works\cite{legeza} it is known, that not only the rung but also the
diagonal and 
plaquette inter-leg couplings are relevant perturbations which generate a 
gap and drive the system into phases with topological order. Depending on 
the sign of the perturbation the emerging phases belong to one of the 
two possible topological sectors.

We have therefore investigated the effect of these couplings by doing calculations
at two specialy chosen values. The finite size scaling analysis 
of the results obtained for couplings $J_{\rm d}=\pm 1$ or $J_{\rm D}=\pm 1$
indicated that irrespective 
whether only one or both diagonal couplings are present, ${\cal O}_{\rm odd} 
\neq 0$ and ${\cal O}_{\rm even} = 0$ for positive $J_{\rm d}$ or $J_{\rm D}$,
while we have ${\cal O}_{\rm even} \neq 0$ and ${\cal O}_{\rm odd} = 0$ for
negative $J_{\rm d}$ or $J_{\rm D}$.

Similarly, calculations for the values $\pm 1$ of the couplings 
$V_{\rm ll}$, $V_{\rm rr}$ and $V_{\rm dd}$ (the other couplings set
to zero) convinced us that the even order parameter is finite, 
${\cal O}_{\rm even}>0$, for $V_{\rm ll}<0$, or $V_{\rm rr}>0$ or
$V_{\rm dd}<0$, while the odd order parameter is finite, ${\cal O}_{\rm odd}>0$,
for opposite signs of the couplings, i.e., for $V_{\rm ll} > 0$, or $V_{\rm rr} < 0$
or $V_{\rm dd} > 0$. 

Our results obtained for the string order parameters allows us
to conclude that the topology of the valence bond structure can be
identified directly by calculating the order parameters ${\cal O}_{\rm odd}$
or ${\cal O}_{\rm even}$, since they mutually exclude each other.

\subsection{The $J_{\rm R}$ - $J_{\rm D}$ phase diagram}

Next we consider the subspace spanned by $J_{\rm R}$ and $J_{\rm D}$.
According to the recent bosonization study\cite{kim} the phase 
transition line separating the two topological sectors is of first order in
the quadrant $J_{\rm R}> 0$, $J_{\rm D} > 0$, while of second order in the 
quadrant $J_{\rm R}< 0$, $J_{\rm D} < 0$. In order to check the behavior
of the string order parameter, we have calculated them along two paths in 
the $(J_{\rm R},J_{\rm D})$ parameter space. 

Path 1 is defined by $J_{\rm R}=1-J_{\rm D}$ with 
$0<J_{\rm D}<1$,  and path 2 is parametrized by 
$J_{\rm R}=-1-J_{\rm D}$ with $-1<J_{\rm D}<0$. They 
are indicated by solid lines on Fig.\ \ref{fig:j1_j23_phase}.

Our numerical results for ${\cal O}_{\rm even}$ and ${\cal O}_{\rm odd}$ 
obtained in the thermodynamic limit along path 1 is plotted on 
Fig.\ \ref{fig:j1_j23}. The lines connecting the points are only guides to the eye.
It is apparent from the figure that ${\cal O}_{\rm even}$ is finite
in one particular region while ${\cal O}_{\rm odd}$ in the other.

As one sees from the figure the string order parameters, extrapolated to 
the infinite ladder limit, have a jump discontinuity at the transition point.
In order to further support the first order nature of the phase
transition we have also investigated the low lying energy spectrum.
We found that there is a sudden change in the degeneracy of the ground state
at the transition point. In the case of open ladders the fourfold degeneracy
of the AKLT phase associated with spin-1/2 end-spins disappear, and the ground
state becomes unique as a rung-dimer-like structure emerges. Moreover,
by examining higher lying energy levels we found that the energy gap
remains finite throughout the transition. The ground state energy shows
a cusp-like singularity, a characteristic feature of first order transitions,
as is shown on the inset of Fig.\ \ref{fig:j1_j23}. Note, however, that for 
finite chains there is no observable level crossing in the ground state due 
to hybridization.

Next we consider the behavior of the order parameters calculated along path 2 
in the negative $J_{\rm R}$ and $J_{\rm D}$ quadrant. Our results are plotted 
in Fig.~\ref{fig:j1_j23_a}. In this case again, only one of the order 
parameters is finite in a particular phase, however, the gap vanishes 
continuously at the phase boundary, which is a characteristic features of 
second order phase transitions. It is worth noting that there is an order of
magnitude difference in the scale of Fig.\ \ref{fig:j1_j23} and
Fig.\ \ref{fig:j1_j23_a} as a function of $J_{\rm D}$.

Since the phase transition points were located on two special 
trajectories only, the phase boundary that separates the two types of 
Haldane-like phases is drawn by dashed line in Fig.\ \ref{fig:j1_j23_phase} 
to indicate that its exact shape was not determined.

\subsection{The role of plaquette couplings}

In order to consider the influence of the plaquette couplings we have chosen
the special values $K\equiv J_{\rm R}= V_{\rm ll}= V_{\rm dd}$, $V_{\rm rr}=0$,
since the phase diagram has already been
studied for this parametrization.\cite{legeza} This choice
has been motivated by the possible mapping of the ladder model
to the spin-1 bilinear-biquadratic chain for $K=1$.\cite{legeza}
Here we calculate
string order parameters along four lines going through the expected 
phase boundaries shown by solid lines in Fig.\ \ref{fig:j145_j23_phase}.

Calculations along path 1 have revealed that ${\cal O}_{\rm odd}$ is finite 
in the AKLT-like Haldane phase, but it vanishes in the dimerized phase. On 
the other hand, ${\cal O}_{\rm even}$ becomes finite in the dimerized phase, 
thus the two phases are well distinguished by the order parameters. Examining 
the trajectory (path 2) going through the phase boundary between the rung 
singlet like Haldane and dimerized phases, we have found that 
${\cal O}_{\rm even}$ remains finite and ${\cal O}_{\rm odd}$ scales to zero
everywhere. This shows that the string order parameter does not uniquely
characterize the Haldane phase, it is rather an indicator of the
topology of the valence bond structure, since both the rung singlet and
the dimerized phases have even topological quantum numbers.

Thus when ${\cal O}_{\rm even} \neq 0$, the string order is not sufficient 
to classify unambiguously the ground state. Another order parameter,
the so-called {\em dimer order} parameter has to be measured. Taking the 
difference of the local energy on neighboring bonds in the middle of the 
ladder this order parameter is defined as
\begin{equation}
  D = \frac{\langle h_{i,i+1} \rangle -
  \langle  h_{i+1,i+2} \rangle}
{\frac{1}{2}\left[ \langle h_{i,i+1} \rangle+
\langle h_{i+1,i+2} \rangle\right] } \,.
\end{equation}
In the Haldane-type spin liquids $D$ vanishes, but there exists an extended
region in the parameter space, where the dimer order parameter
scales to finite value in the thermodynamic limit, signaling a
dimerized phase with spontaneously broken translational symmetry.

Next let us consider the $K>0$ half-plane focusing on the $J_{\rm D}=0$
half-line and its vicinity. Calculation of the order parameters
along path 3, defined by $K=1+J_{\rm D}$ with $-1<J_{\rm D}<1$ 
revealed that ${\cal O}_{\rm even}$ is finite for $J_{\rm D}<0$,
while ${\cal O}_{\rm odd}$ is finite for $J_{\rm D}>0$. At $J_{\rm D}=0$ all
order parameters including $D$ vanish as is shown in Fig.\ \ref{fig:j1_j2_j3}.
Since the string order parameters become finite for arbitrary small finite value of
$J_{\rm D}$ the two types of spin liquids are separated by a critical
line and the extended critical phase reported earlier\cite{legeza} does not exist.

Finally, it can be seen on the figure that for $J_{\rm D}>0.6$ both order 
parameters vanish and a new phase appears. In order to 
analyze the question whether the new phase develops just above the $VBS$ 
point, we investigated the vicinity of the $VBS$ point along path 4. We 
found that ${\cal O}_{\rm odd}$ remains finite up to $K\sim 1.9$ while 
${\cal O}_{\rm even}$ vanishes for the whole regime. Investigating the 
low lying energy spectra we found a multiply degenerate ground state. In 
order to decide whether this phase extends to the $J_{\rm D}$ axis we made 
test calculations at $J_{\rm D}=0.05$, $K=1.5, 2$ and found a finite 
${\cal O}_{\rm odd}$ but vanishing ${\cal O}_{\rm even}$ order parameters. 
This suggests the schematic phase boundary shown 
by a dashed line in the figure.

\section{Conclusions}

In the present paper we have studied the relationship between various gapped 
Haldane and non-Haldane spin liquid phases and the two different string order 
parameters in the general two-leg spin ladder model, containing rung, diagonal 
and plaquette couplings as well. The string order parameters have been 
calculated numerically using the DMRG method. 

It has been found that there are two kinds of Haldane-type massive spin liquids
characterized by finite $O_{\rm even}$ or $O_{\rm odd}$ string order 
parameters, respectively. It was shown that ${\cal O}_{\rm odd}$ is
finite in the AKLT-like phase, while ${\cal O}_{\rm even}$ is finite
in the rung singlet like phase. Examining the transition from the AKLT-like 
to the rung-singlet-like phases the type of phase transition can be first 
or second order depending on the path in parameter space, and this shows up
in the abrupt or continuous vanishing of the order parameter.

When plaquette couplings are allowed for, non-Haldane-like spin liquid
phases can occur, and the string order parameters are not sufficient
to characterize the phases. ${\cal O}_{\rm even}$ remains finite
in the dimerized phases, too. Thus the string order parameter reflects
the topology of the valence bond structure and does not characterize
uniquely the Haldane-type phases.
A dimer order parameter can, however, be defined, which is finite in a
non-Haldane-type dimer phases but vanishes in Haldane-type spin liquids.

It was found previously that there exists a small range of the parameters where
the combination of perturbing operators is irrelevant and the spectrum remains
massless. Investigating the vicinity of $J_{\rm D}=0, K>0$ axis we have
found that both string order parameters vanish on this critical line, but
${\cal O}_{\rm even}$ or ${\cal O}_{\rm odd}$ becomes finite immediately as we move 
away from this axis. Therefore, we conclude that the two kind of spin liquids 
are separated by a critical line and not by an extended gapless regime 
reported earlier. In addition, around the $J_{\rm D}=1$ line for $K>1.8$ 
we have found a new phase. This phase is characterized by multiply 
degenerate ground state and vanishing order parameters. The detailed analysis of
this region is left for future investigation. 

\acknowledgments

This research was supported in part by the Hungarian Research Fund
(OTKA) Grant No.\ T30173, F31949 and F32231. G.\ F.\ was also supported
by the Bolyai Research Scholarship. \"O. L. acknowledges the hospitality
of the Friedrich Alexander University of N\"urnberg where part of
the work was completed.

\newpage

\begin{widetext}

\begin{figure}[bp]
\includegraphics[scale=.5]{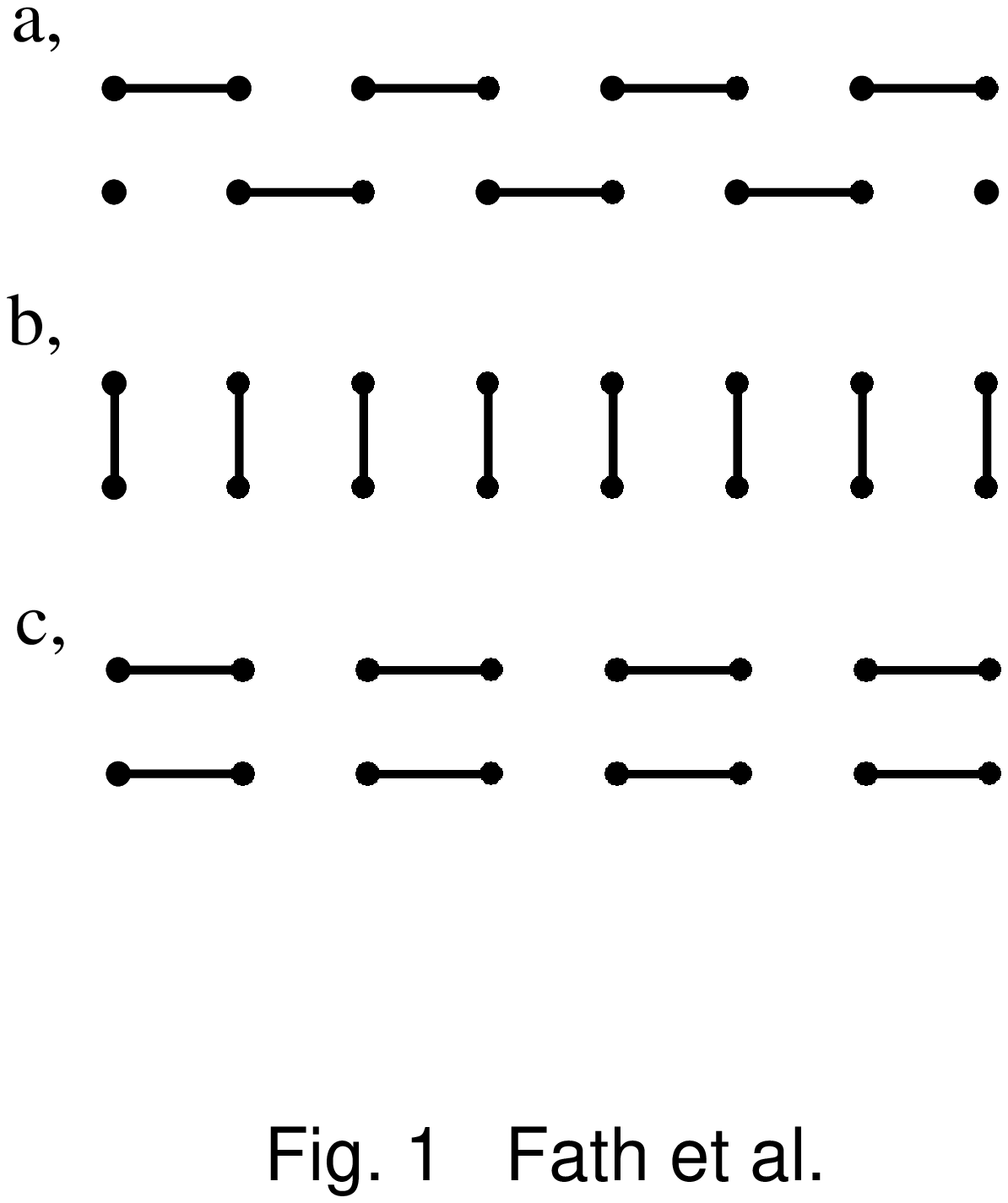}
\caption{Representative valence-bond configurations without quantum 
fluctuations in different gapped phases of spin ladders. (a) The 
AKLT-like state, (b) rung singlet state, (c) dimerized state. State 
(a) is odd, while states (b) and (c) are even under our
topological classification.}
\label{fig:topo}
\end{figure}

\begin{figure}[tbp]
\includegraphics[scale=.5]{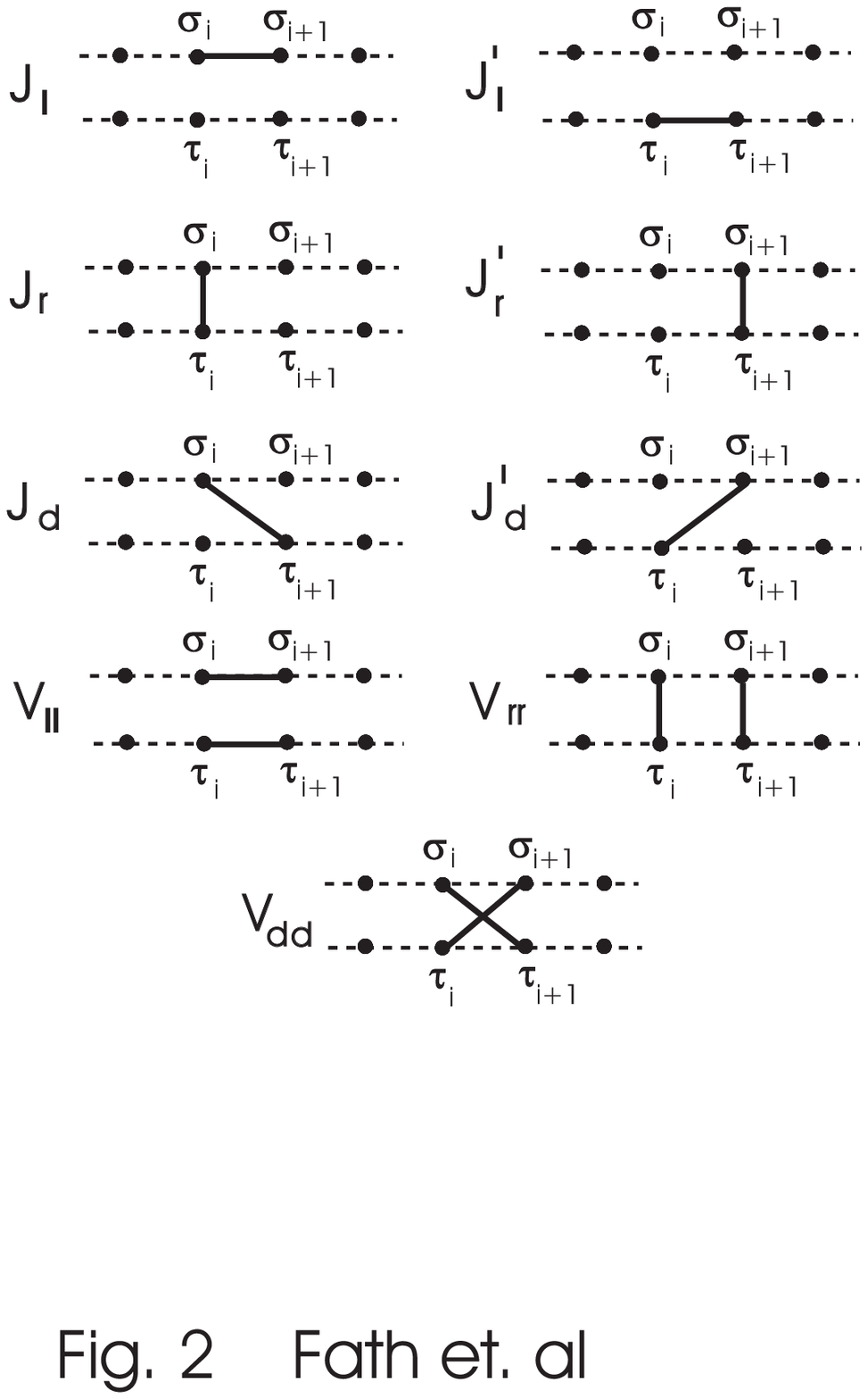}
\caption{Schematic plot of the Hamiltonian}
\label{fig:ham}
\end{figure}

\begin{figure}[tbp]
\includegraphics[scale=.5]{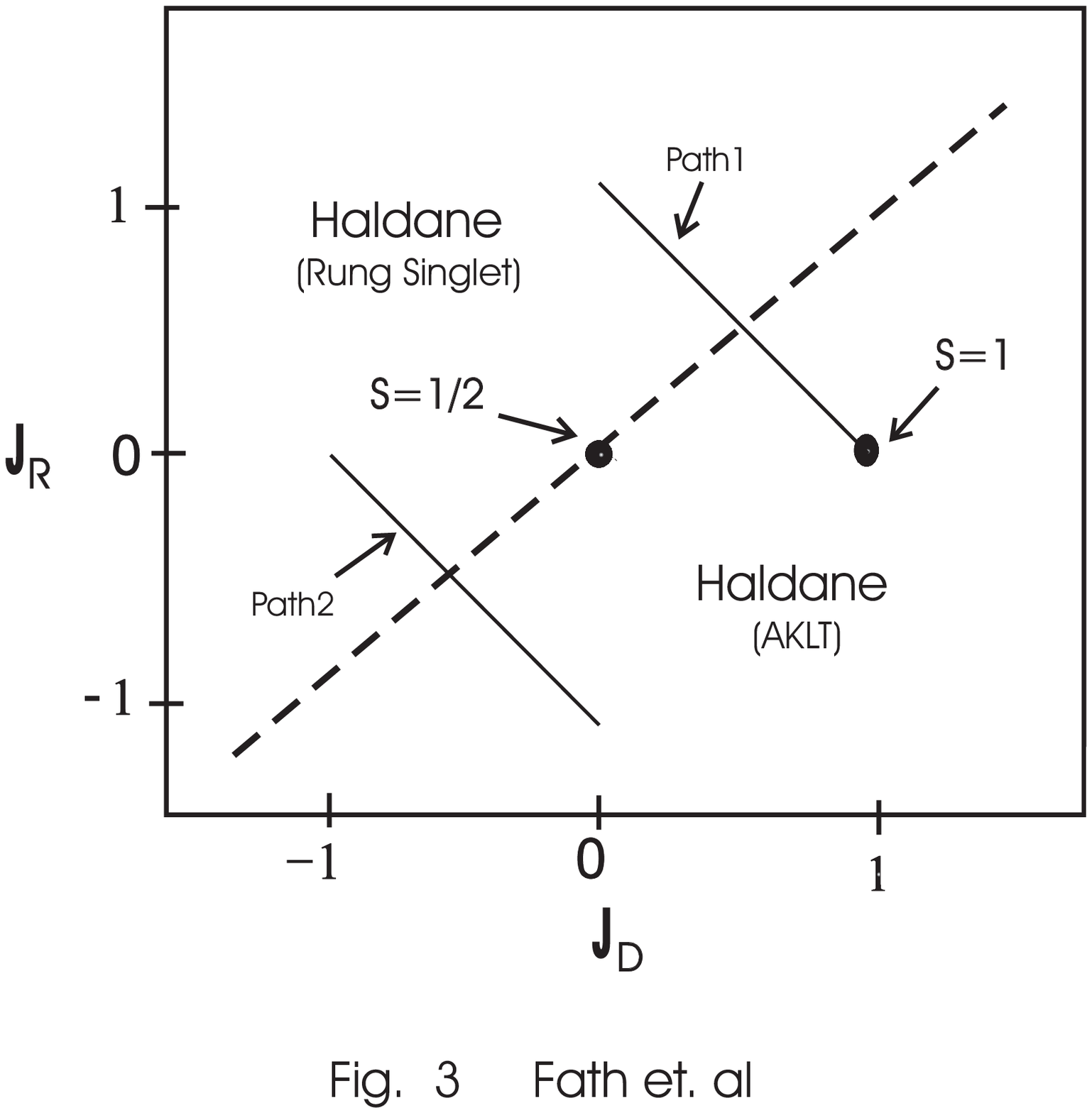}
\caption{The $J_{\rm R}$-$J_{\rm D}$ phase diagram. Phase boundary is 
indicated by dashed line and the two trajectories of the 
calculation by solid lines.}\label{fig:j1_j23_phase}
\end{figure}

\begin{figure}[tbp]
\includegraphics[scale=.5]{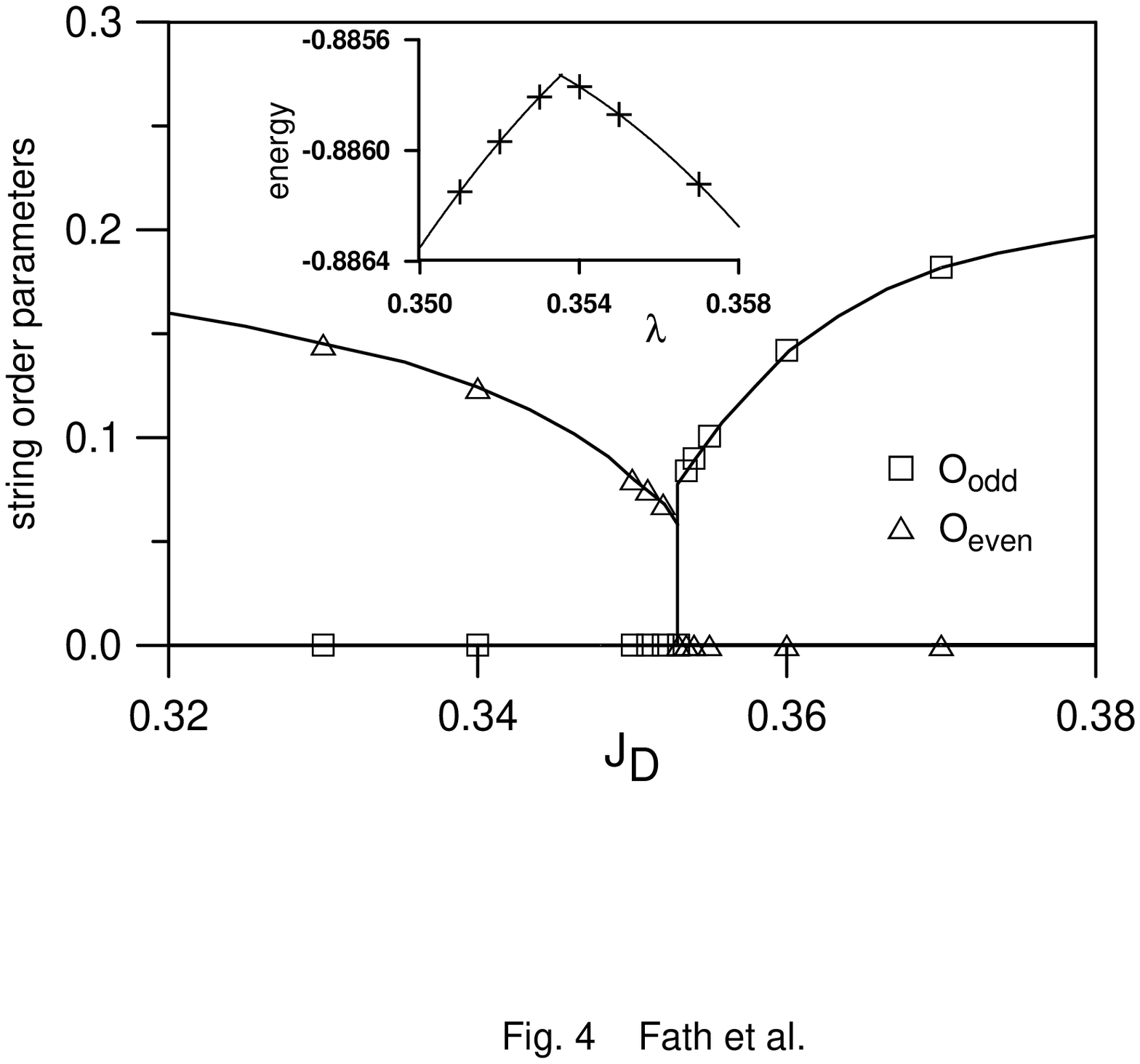}
\caption{The $N\to\infty$ extrapolated values of the string
order parameters ${\cal O}_{\rm even}$ and ${\cal O}_{\rm odd}$
calculated along path 1. The inset of the figure shows
the cusp singularity of the ground state energy indicating a first
order transition. }\label{fig:j1_j23}
\end{figure}

\begin{figure}[tbp]
\includegraphics[scale=.5]{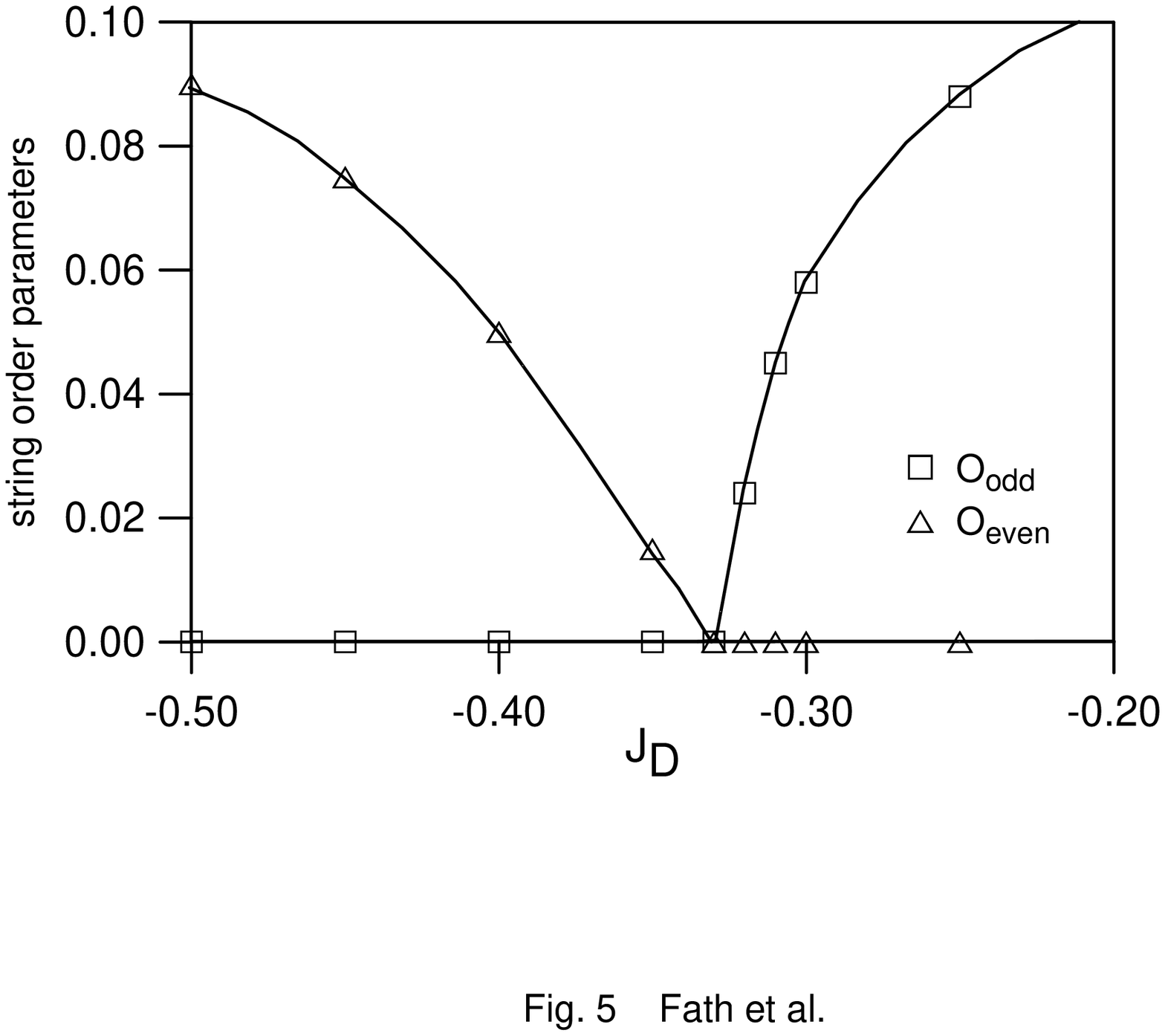}
\caption{The ${\cal O}_{\rm even}$ and ${\cal O}_{\rm odd}$ string order
parameters calculated along path 2.}\label{fig:j1_j23_a}
\end{figure}

\begin{figure}[tbp]
\includegraphics[scale=.5]{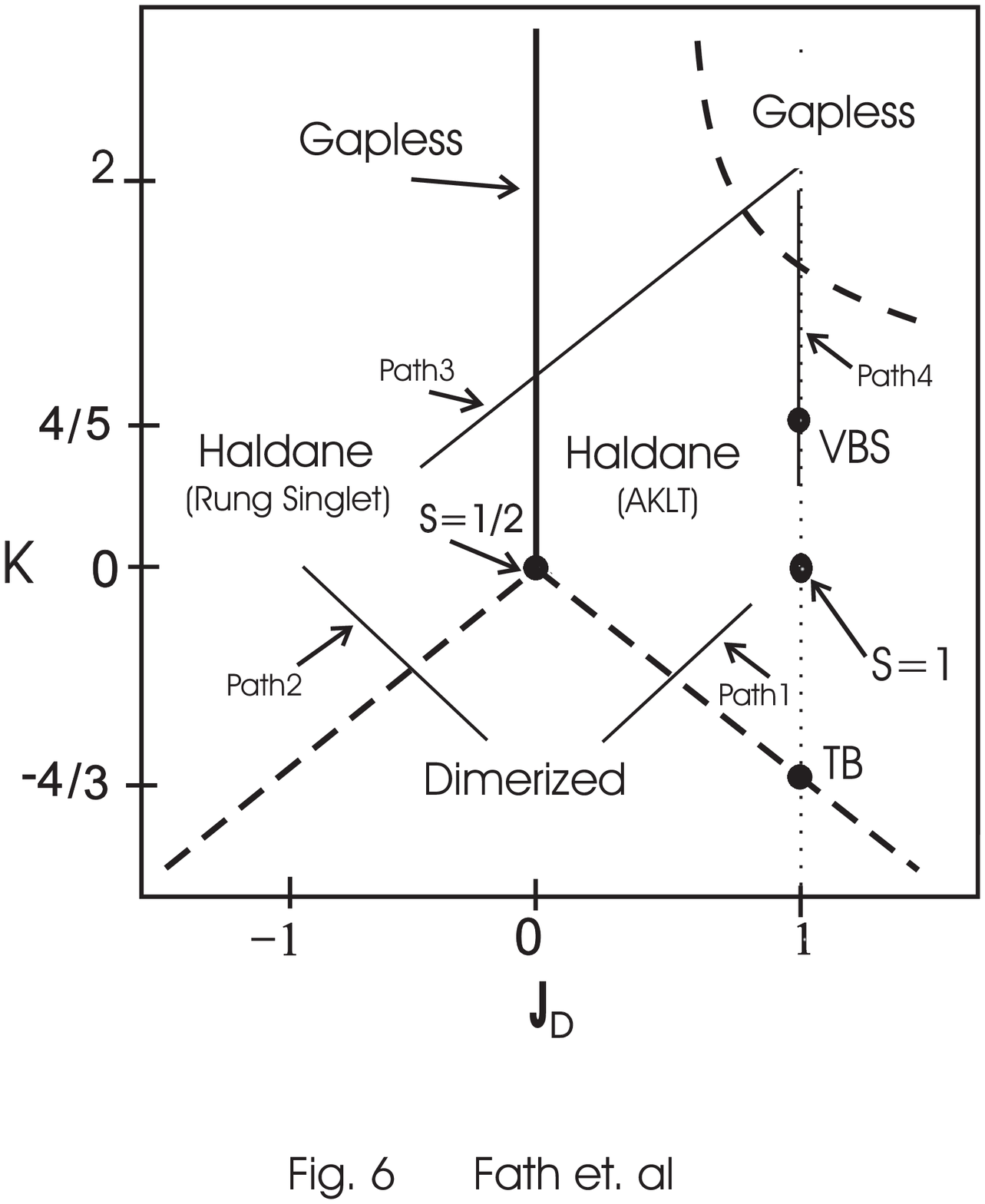}
\caption{The $K$ vs. $J_{\rm D}$ phase diagram. Phase boundaries are
indicated by dashed lines and the various trajectories of the calculation by 
solid lines.}
\label{fig:j145_j23_phase}
\end{figure}

\begin{figure}[tbp]
\includegraphics[scale=.5]{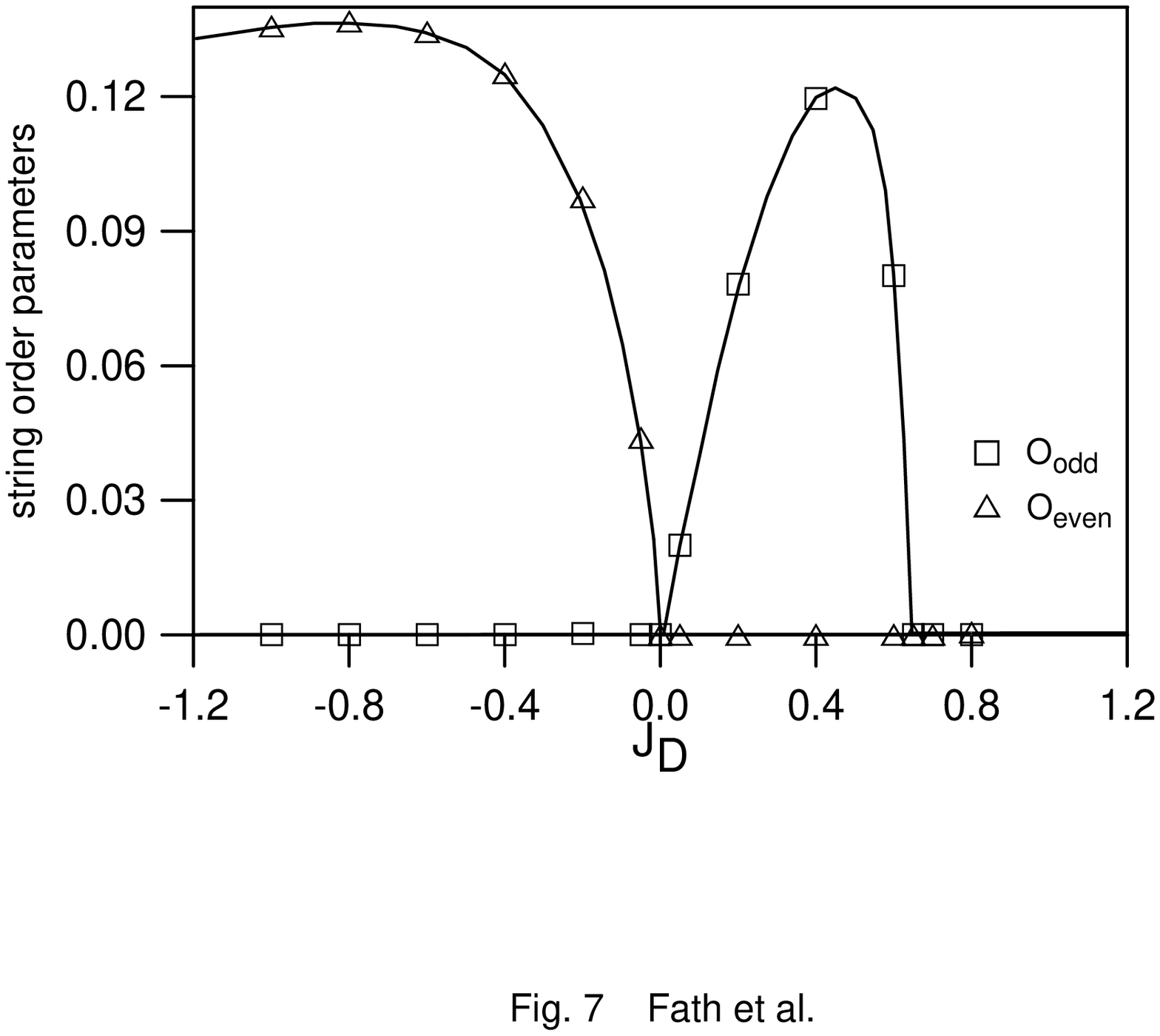}
\caption{The ${\cal O}_{\rm even}$ and ${\cal O}_{\rm odd}$ string order
 parameters calculated along path 3.}
\label{fig:j1_j2_j3}
\end{figure}

\end{widetext}


\begin{references}
 
\bibitem{review}{For a review see E. Dagotto, cond-mat/9908250.}

\bibitem{ferro}{K.\ Hida, J.\ Phys.\ Soc.\ Jpn.\ {\bf 60}, 1347 (1991);
        H.\ Watanabe, K.\ Nomura, and S.\ Takada,
        J.\ Phys.\ Soc.\ Jpn.\ {\bf 62}, 2845 (1993):
        H.\ Watanabe, Phys.\ Rev.\ B {\bf 50}, 13442 (1994).}

\bibitem{haldane}{F.\ D.\ M.\ Haldane,
         Phys.\ Rev.\ Lett.\ {\bf 50}, 1153 (1983);
         Phys.\ Lett.\ {\bf 93A}, 464 (1983).}

\bibitem{AKLT}{I.\ Affleck, T.\ Kennedy, E.\ Lieb, and H.\ Tasaki,
         Phys.\ Rev.\ Lett.\ {\bf 59}, 799 (1987);
         Commun.\ Math.\ Phys.\ {\bf 115}, 477 (1988).}
         
\bibitem{antiferro}{E.\ Dagotto, J.\ Riera, and D.\ J.\ Scalapino,
         Phys.\ Rev.\ B {\bf 45}, 5744 (1992);
         T.\ Barnes, E.\ Dagotto, J.\ Riera, and E.\ S.\ Swanson, 
         Phys.\ Rev.\ B {\bf 47}, 3196 (1993);
         S.\ R.\ White, R.\ M.\ Noack, and D.\ J.\ Scalapino,
         Phys.\ Rev.\ Lett.\ {\bf 73}, 886 (1994).}

\bibitem{legeza}{\"O. Legeza, G.\ F\'ath, and J.\ S\'olyom,
         Phys.\ Rev.\ B {\bf 55}, 291 (1997).} 

\bibitem{nersesyan}{A.\ A.\ Nersesyan and A.\ M.\ Tsvelik,
         Phys.\ Rev.\ Lett.\ {\bf 78}, 3939 (1997).}

\bibitem{white1}{S.\ R.\ White,
         Phys.\ Rev.\ B {\bf 53}, 52 (1996).}

\bibitem{kim}{E.\  Kim, G.\ F\'ath, J.\ S\'olyom and J.\ Scalapino,
         cond-mat/9910023.} 

\bibitem{bonesteel}{N.\ E,\ Bonesteel, 
         Phys.\ Rev.\ B {\bf 40}, 8954 (1989).}

\bibitem{nedelcu}{C.-M.\ Nedelcu, A.\ K.\ Kolezhuk and 
         H.-J.\ Mikeska, J.\ Phys.: Condens.\ Matter {\bf 12}, 959 (2000).}

\bibitem{dennijs}{M.\ P.\ M.\ den Nijs and K. Rommelse, 
         Phys.\ Rev.\ B {\bf 40}, 4709 (1989).}

\bibitem{dmrg}{S.\ R.\ White,
         Phys.\ Rev.\ Lett.\ {\bf 69}, 2863 (1992);
         Phys.\ Rev.\ B {\bf 48}, 10345 (1993).}

\bibitem{kolezhuk}{A.\ K.\ Kolezhuk and H.-J.\ Mikeska,
         Int.\ J.\ Mod.\ Phys.\ B {\bf 12}, 2325 (1998).
         \label{kolezhuk}}

\bibitem{nishi}{Y.\ Nishiyama, N.\ Hatano, and M.\ Suzuki,
         J.\ Phys.\ Soc.\ Jap.\ {\bf 64}, 1967 (1995).\label{nishi}}

\end{references}
\end{document}